\documentclass[a4paper,11pt]{article}

\usepackage[margin=1in]{geometry}
\usepackage[T1]{fontenc}
\usepackage{newtxtext}
\usepackage{microtype}

\usepackage{amsmath}
\usepackage{amsthm}
\usepackage{mathtools}
\usepackage{bm}
\usepackage{newtxmath}
\usepackage{mathrsfs}

\usepackage{graphicx}
\usepackage{booktabs}
\usepackage{enumitem}

\usepackage[numbers,sort&compress]{natbib}
\usepackage{xcolor}
\usepackage[
  colorlinks=true,
  linkcolor=blue!55!black,
  citecolor=blue!55!black,
  urlcolor=blue!55!black
]{hyperref}

\newtheorem{theorem}{Theorem}[section]

\theoremstyle{definition}

\title{\bfseries A Sklar-Type Decomposition for Positive Binary Distributions and a Gibbs Interpretation}

\author{
Yi-Yu Lin$^{1,2,*}$\\[0.5em]
\small $^{1}$Fudan Center for Mathematics and Interdisciplinary Study,
Fudan University, Shanghai 200433, China\\[0.4em]
\small $^{2}$Shanghai Institute for Mathematics and Interdisciplinary Sciences
(SIMIS), Shanghai 200433, China\\[0.5em]
\small $^{*}$\texttt{yiyu@simis.com}
}

\date{}

\begin{document}

\maketitle

\begin{abstract}

We study strictly positive multivariate binary distributions by separating one-body marginals from dependence. Local positive reweighting defines a dependence nucleus, completely characterized by the higher-order Walsh log-linear coefficients $J_S$, $|S|\ge2$. Any such nucleus and any strictly positive one-body marginals determine a unique joint distribution. In finite-spin Gibbs systems, these coefficients are the many-body interaction couplings.
\end{abstract}

\vspace{0.5em}

\noindent

\tableofcontents

\section{Introduction}
\label{sec:introduction}

Discrete probability distributions are ubiquitous in physics. For example,
finite-spin systems in classical statistical mechanics, as well as the
measurement statistics obtained by applying the Born rule to multiqubit
quantum states, naturally give rise to discrete joint probability
distributions. For a joint probability distribution, a basic structural
question is how to distinguish the statistical laws of the individual
variables from the relations among them---and this is precisely the question
answered by the celebrated theorem of Sklar~\cite{sklar1959}.

Sklar's theorem tells us that a continuous joint distribution can be uniquely
decomposed into two parts: how the individual variables are distributed, and
how they are jointly distributed once expressed in their respective
percentile coordinates. Conversely, any collection of continuous marginals,
together with any such dependence structure, determines a unique joint
distribution. Schematically,
\begin{equation}
    F
    \longleftrightarrow
    \bigl(\{F_i\},\, C\bigr).
\end{equation}
The cumulative distribution function of this joint distribution of percentile
variables, whose one-body marginals are uniform, is traditionally called the
\emph{copula}.

The work of Geenens~\cite{geenens2020,geenens2025} suggests a discrete
structure with the same Sklar spirit. A strictly positive binary joint
probability table can be uniquely decomposed into two parts: the one-body
marginals of the individual variables, and a relational structure that remains
unchanged under local positive reweighting. Conversely, any such relational
structure, together with any collection of strictly positive one-body
marginals, determines a unique joint probability table. Schematically,
\begin{equation}
    p
    \longleftrightarrow
    \bigl(\{p_i\},\,[p]\bigr).
\end{equation}

Here $[p]$ denotes an entire family of probability tables related to one
another by local reweighting. It expresses the idea that we allow the
probabilistic tendencies of the individual parties to change, while regarding
the way in which they are combined as unchanged. It is therefore an abstract
relational object, referred to in Ref.~\cite{geenens2025} as a
\emph{dependence nucleus}. The continuous counterpart of $[p]$ is the class
of joint distributions sharing the same copula. In particular, if
$\gamma_p$ denotes the unique joint probability table in $[p]$ with uniform
one-body marginals, then the joint cumulative distribution function obtained
by summing $\gamma_p$ may naturally be regarded as playing the role of a
discrete copula.

Clearly, the continuous and discrete pictures share the same beautiful idea:
the statistical data of a whole composed of several parts can be elegantly
separated into the statistics of the individual parts and the statistical data
that characterize the relations among them, while the two together are
sufficient to recover the whole uniquely.

Notice that our concern here is not to decide ``which function on
$[0,1]^N$ should be called the discrete copula.'' Rather, we remain entirely
within the finite probability table itself and seek a strict decomposition:
\begin{equation}
    \text{joint probability table}
    \quad\longleftrightarrow\quad
    \bigl(
        \text{one-body marginals},
        \text{margin-free relational structure}
    \bigr).
\end{equation}
More explicitly, we shall prove step by step that such a relational structure
indeed exists, that it can be characterized completely, and that once it is
recombined with arbitrary admissible marginals, it recovers exactly one joint
table. The overall route is summarized in Table~\ref{tab:roadmap}.

\begin{table}[t]
    \centering
    \begin{tabular}{p{0.43\textwidth} p{0.48\textwidth}}
        \toprule
        Question & Our answer \\
        \midrule
        What does it mean to change only the individual parties?
        & Local positive reweighting \\[0.4em]

        What remains invariant under such a transformation?
        & All second- and higher-order $J_S$ \\[0.4em]

        Are these invariants complete?
        & Yes. Two tables have the same higher-order $J_S$ if and only if
          they belong to the same local-reweighting equivalence class \\[0.4em]

        Can the relational data and the marginals recover the whole?
        & Yes, and the recovery is unique \\[0.4em]

        Is there a canonical representative?
        & Every equivalence class contains a unique table with uniform
          one-body marginals \\
        \bottomrule
    \end{tabular}
    \caption{The structure developed in this paper.}
    \label{tab:roadmap}
\end{table}

Before entering the main development, we would like to comment fairly on the
history preceding it. As early as 1900, Yule~\cite{yule1900} observed that,
after removing the two marginals of a discrete $2\times2$ table, only one
degree of association remains, which can be represented by the odds ratio and
which later became a standard parameterization in the theory of binary
contingency tables. In 1912, Yule~\cite{yule1912} further proposed using
row--column scaling---what we shall call local reweighting---to standardize
tables with different marginals to tables with common uniform marginals, and
then to compare their association.\footnote{For the classical development of
association measures for contingency tables, see also
Ref.~\cite{goodmanKruskal1954}.}
Remarkably, this predates Sklar's theorem of 1959~\cite{sklar1959}.
Going one step further, Mosteller~\cite{mosteller1968} explicitly proposed
that contingency tables related by row--column scaling form an equivalence
class with the same association, and used the term \emph{basic nucleus}. He
also explicitly advocated choosing a table with uniform marginals from this
equivalence class in order to obtain a clearer view of its association.

The key contribution of Geenens and collaborators
\cite{geenens2020,geenens2025} was to recognize explicitly that these older
ideas from the classical theory of contingency tables already contain the
Sklar spirit needed in the discrete setting. Once the entire equivalence class
generated by marginal transformations is interpreted as a dependence nucleus,
there is no need to regard a discrete copula as an arbitrarily chosen
continuous extension. Instead, the joint probability mass function with
uniform marginals obtained through row--column reweighting can be regarded as
a \emph{copula pmf}, or \emph{copula array}, playing in the discrete setting
the role played by the continuous copula.

What we especially wish to emphasize is that, along the line of thought in
which probability tables differing only by local reweightings of the
individual parties are regarded as having the same dependence, the
corresponding theory for general finite multivariate discrete distributions
has already been developed in a systematic and rather complete way.
First, Ref.~\cite{geenens2025} shows that for finite-dimensional discrete
joint distributions a unified, general, and computationally accessible
multivariate discrete-copula framework can be constructed through the idea of
an $I$-projection: among all candidate probability arrays with uniform
marginals, one chooses the array closest to the original table in the
Kullback--Leibler sense. On the other hand, results from the theory of
multivariate contingency tables show that one-body marginals and suitably
chosen conditional odds-ratio parameters can be specified as mutually
independent parameter blocks and together uniquely determine a strictly
positive joint distribution; this result has also been reformulated directly
in the language of discrete copulas
\cite{bergsmaRudas2002Variation,rudas2018,liu2025DiscreteCopulaDiffusion}.

We therefore do not seek to add further generality to this existing theory.
In a sense, our interest lies in precisely the opposite direction: we return
to a special but structurally unusually transparent sector---the strictly
positive $N$-binary case---and write the comparatively abstract discrete
Sklar structure of the general theory in a completely explicit form.
For a joint distribution on $x_i\in\{-1,+1\}$,
the logarithm of the probability can be expanded uniquely as
\begin{equation}
    \log p(x)
    =
    \sum_{S\subseteq[N]} J_S\,\chi_S(x),
    \qquad
    \chi_S(x)
    =
    \prod_{i\in S}x_i .
    \label{eq:intro-walsh}
\end{equation}
Under a local reweighting,
\begin{equation}
    p(x)
    \longmapsto
    \frac{
        p(x)\prod_i f_i(x_i)
    }{
        \sum_y p(y)\prod_i f_i(y_i)
    },
    \label{eq:intro-reweighting}
\end{equation}
only the constant term and the one-body coefficients $J_i$ can change, while
all coefficients involving at least two variables, $J_S$, $|S|\geq2$, remain unchanged. Conversely, if two strictly positive binary tables have
exactly the same higher-order coefficients, then they necessarily differ only
by such a local reweighting. Hence,
\begin{equation}
    [p]
    \longleftrightarrow
    \{J_S:\ |S|\geq2\},
    \label{eq:intro-nucleus-coordinates}
\end{equation}
that is, an abstract dependence structure can be represented completely and
explicitly by the full collection of higher-order $J_S$.

The significance of this representation is not merely that it provides a
convenient set of coordinates. The same $J_S$ already have familiar meanings
in different theories. In the theory of contingency tables they are the
Walsh log-linear interactions, and in particular they can be read
equivalently from conditional odds ratios. More interestingly, the same
framework admits a direct interpretation in statistical mechanics: in a
finite-spin Gibbs system, the higher-order $J_S$ correspond precisely to
two-body and higher-order interaction couplings. Thus, in this explicit
binary setting, the corresponding objects appearing in discrete copula
theory, log-linear theory, and the many-body couplings of classical
statistical mechanics are brought into a common description.

The paper is organized as follows. Section 2 reviews Sklar's theorem, introduces the positive binary setting, and uses the $2\times2$ table as a warmup. Section 3 develops the multipartite log-linear description and identifies the higher-order Walsh coefficients as complete coordinates of the dependence nucleus. Section 4 proves the positive binary Sklar-type decomposition, namely that the dependence nucleus and the one-body marginals uniquely determine the joint distribution. Section 5 discusses the Gibbs interpretation and concludes.

\section{Background and Warmup}
\label{sec:background}

\subsection{Review of Sklar's theorem}
\label{subsec:sklar-review}

The standard statement of Sklar's theorem~\cite{sklar1959} is the
following. Let $F(x_1,\ldots,x_N)$
be an $N$-dimensional joint cumulative distribution function (CDF),
with marginal distribution functions $F_1,\ldots,F_N$.
Then there exists an $N$-dimensional copula $C$ such that
\begin{equation}
    F(x_1,\ldots,x_N)
    =
    C\bigl(F_1(x_1),\ldots,F_N(x_N)\bigr).
    \label{eq:sklar}
\end{equation}
If the marginal distributions are continuous, then $C$ is unique.
\footnote{More generally, $C$ is uniquely determined on $    \operatorname{Ran}(F_1)\times\cdots\times\operatorname{Ran}(F_N)$.}
Conversely, given any collection of one-variable distribution functions
$F_1,\ldots,F_N$ and any $N$-dimensional copula $C$,
Eq.~\eqref{eq:sklar} defines a legitimate joint CDF with marginals $F_i$.

For a continuous random variable $X_i$, the quantity
\begin{equation}
    u_i
    =
    F_i(x_i)
    =
    \Pr(X_i\leq x_i),
    \qquad
    i=1,\ldots,N,
\end{equation}
has a simple intuitive meaning: it is the fraction of the marginal
distribution of $X_i$ lying at or below $x_i$.
In this sense, $u_i$ is the percentile coordinate of $x_i$.
The copula can therefore also be viewed as the joint CDF of a set
of random variables $U_i$:
\begin{equation}
    C(u_1,\ldots,u_N)
    =
    \Pr(U_1\leq u_1,\ldots,U_N\leq u_N),
\end{equation}
where each one-variable marginal $U_i$ is uniformly distributed on
$[0,1]$.

In principle, a discrete joint distribution can of course still be
written in the form of Eq.~\eqref{eq:sklar}. The difficulty is that the
CDFs of discrete marginals are step functions, and hence do not sweep
through the whole interval $[0,1]$.
\footnote{For example, a binary variable may have $    \operatorname{Ran}(F_i)=\{0,p_i,1\}$.
}
The joint distribution therefore determines $C$ only on the discrete
grid formed by the ranges of the marginal CDFs, and does not uniquely
determine what happens between the grid points. In other words, more than
one continuous copula extension can be combined with the same discrete
marginals to produce exactly the same discrete joint distribution.
The key insight of Refs.~\cite{geenens2020,geenens2025} is that, rather
than arbitrarily choosing one among these non-unique continuous copula
extensions, one should separate three levels that nearly coincide
automatically in the continuous case. One first defines what it means
to have the same dependence, $    [p]$,
then chooses a canonical table within this class, $  \gamma_p$,
and, if one wishes to construct an object of exactly the same mathematical
type as a continuous copula, one finally takes its cumulative distribution
function, $    C_p$.

\subsection{Setup}
\label{subsec:setup}

Consider $N$ binary variables
\begin{equation}
    X_i\in\{-1,+1\},
    \qquad
    i=1,\ldots,N.
\end{equation}
A joint probability table consists of $2^N$ numbers
\begin{equation}
    p(x_1,\ldots,x_N)>0,
    \qquad
    \sum_{x_1,\ldots,x_N}
    p(x_1,\ldots,x_N)=1.
    \label{eq:positive-binary-table}
\end{equation}
Throughout this paper, we restrict attention to the case in which every
cell probability is strictly positive.
\footnote{This assumption can be relaxed, but doing so introduces technical
issues that are not our main concern here.}
The marginal distribution of the $i$th variable is
\begin{equation}
    p_i(s)
    =
    \sum_{\{x_j:\,j\neq i\}}
    p(x_1,\ldots,x_{i-1},s,x_{i+1},\ldots,x_N).
\end{equation}
Since $X_i$ has only two possible outcomes, its marginal distribution
contains only one independent number:
\begin{equation}
    r_i
    =
    \Pr(X_i=+1),
    \qquad
    \Pr(X_i=-1)=1-r_i.
\end{equation}
Thus, all one-body marginals together account for $N$ degrees of freedom.

The full joint table, on the other hand, has $    2^N-1$ degrees of freedom, because the $2^N$ probabilities must sum to one.
This already suggests that, after removing the $N$ one-body degrees of
freedom, the relational part should contain $    2^N-N-1$
degrees of freedom. 
But counting degrees of freedom tells us only how much ``should remain''.
It does not yet tell us:
\begin{enumerate}
    \item which quantities genuinely represent the relations;
    \item whether these quantities are complete;
    \item whether they can be freely and uniquely recombined with the
    marginals.
\end{enumerate}
These are the questions that we shall actually prove.

\subsection{Warmup: the smallest $2\times2$ probability table}
\label{subsec:warmup}

Let $    X,Y\in\{-1,+1\}$.
The joint probability table can be written as
\begin{equation}
\begin{array}{c|cc}
    & Y=+1 & Y=-1 \\ \hline
X=+1 & p_{++} & p_{+-} \\
X=-1 & p_{-+} & p_{--}
\end{array}.
\end{equation}
There are four cells, but normalization removes one degree of freedom,
so the table has three independent degrees of freedom.
Each of the two one-body marginals contributes one degree of freedom.
The purely relational part should therefore contain only one remaining
degree of freedom. Classical contingency-table theory tells us that this
degree of freedom can be represented by the odds ratio~\cite{yule1900,yule1912}
\begin{equation}
    \omega(p)
    =
    \frac{p_{++}p_{--}}
         {p_{+-}p_{-+}}.
    \label{eq:odds-ratio}
\end{equation}

Let us review the relevant properties of the odds ratio in some detail.
We first introduce the notion of local reweighting. Assign positive
weights to the two outcomes of $X$ and independently to the two outcomes
of $Y$:
\begin{equation}
    a_+,a_->0,
    \qquad
    b_+,b_->0.
\end{equation}
We then define a new table by
\begin{equation}
    q_{st}
    =
    \frac{a_s b_t\,p_{st}}{Z},
    \qquad
    Z
    =
    \sum_{s,t}a_s b_t\,p_{st}.
    \label{eq:2x2-reweighting}
\end{equation}
The factor $Z$ simply ensures that the entries of the new table again sum
to one.
This transformation has a very simple interpretation:
\begin{itemize}
    \item $a_s$ changes the relative weights of the two rows;
    \item $b_t$ changes the relative weights of the two columns;
    \item no irreducible factor $g(s,t)$ depending jointly on both $s$ and
    $t$ has been introduced.
\end{itemize}
In this simple sense, the transformation ``changes only the marginals''.
At this stage, however, we are not yet entitled to declare that it
``changes the marginals but not the relation''. For the moment, we merely
make the following candidate definition:
\begin{equation}
    p\sim q
\end{equation}
if and only if $p$ and $q$ can be transformed into one another by such
positive row--column reweightings.
All probability tables related in this way form an equivalence class,
which we denote by $    [p]$.
Later we shall show that this equivalence class indeed has precisely the
properties required by the Sklar picture. Only then will we have a full
justification for calling it a \emph{dependence nucleus}.

We first show that the odds ratio is invariant under reweighting.
The reason is simple: the ratio is constructed so that the factors applied
separately to the rows and columns cancel exactly.
Using Eq.~\eqref{eq:2x2-reweighting},
\begin{equation}
    \omega(q)
    =
    \frac{q_{++}q_{--}}
         {q_{+-}q_{-+}}.
\end{equation}
Substituting explicitly,
\begin{equation}
    \omega(q)
    =
    \frac{
        (a_+b_+p_{++}/Z)
        (a_-b_-p_{--}/Z)
    }{
        (a_+b_-p_{+-}/Z)
        (a_-b_+p_{-+}/Z)
    }.
\end{equation}
All weights and normalization factors cancel, giving
\begin{equation}
    \omega(q)=\omega(p).
\end{equation}
Thus, all tables in the same reweighting equivalence class necessarily
have the same odds ratio. More importantly, the converse is also true.
Suppose that two strictly positive tables $p$ and $q$ have the same odds
ratio. Define the cellwise ratios
\begin{equation}
    r_{st}
    =
    \frac{q_{st}}{p_{st}}.
\end{equation}
Equality of the odds ratios implies
\begin{equation}
    r_{++}r_{--}
    =
    r_{+-}r_{-+}.
\end{equation}
This condition guarantees that $r_{st}$ factorizes as
\begin{equation}
    r_{st}=a_s b_t.
\end{equation}
Hence
\begin{equation}
    q_{st}=a_s b_t\,p_{st},
\end{equation}
and, once normalization is written explicitly, this is precisely the
local reweighting of Eq.~\eqref{eq:2x2-reweighting}. Therefore, for
strictly positive $2\times2$ tables,
\begin{equation}
    \boxed{
        p\sim q
        \quad\Longleftrightarrow\quad
        \omega(p)=\omega(q)
    }.
    \label{eq:2x2-complete-invariant}
\end{equation}
In other words, the odds ratio is an invariant that completely labels the
entire relational equivalence class.

As a concrete example, consider
\begin{equation}
p=
\begin{array}{c|cc}
    & Y=+1 & Y=-1 \\ \hline
X=+1 & 4/9 & 2/9 \\
X=-1 & 1/9 & 2/9
\end{array}.
\end{equation}
Its marginals are
\begin{equation}
    p_X
    =
    \left(
        \frac{2}{3},
        \frac{1}{3}
    \right),
    \qquad
    p_Y
    =
    \left(
        \frac{5}{9},
        \frac{4}{9}
    \right).
\end{equation}
Its odds ratio is
\begin{equation}
    \omega(p)
    =
    \frac{(4/9)(2/9)}
         {(2/9)(1/9)}
    =
    4.
\end{equation}
Now reweight the two rows by
\begin{equation}
    a_+
    =
    \frac{3}{4},
    \qquad
    a_-
    =
    \frac{3}{2},
    \qquad
    b_+=b_-=1.
\end{equation}
This gives
\begin{equation}
\gamma_p=
\begin{array}{c|cc}
    & Y=+1 & Y=-1 \\ \hline
X=+1 & 1/3 & 1/6 \\
X=-1 & 1/6 & 1/3
\end{array}.
\end{equation}
Both of its marginals are now
\begin{equation}
    \left(
        \frac{1}{2},
        \frac{1}{2}
    \right),
\end{equation}
while its odds ratio remains
\begin{equation}
    \omega(\gamma_p)
    =
    \frac{(1/3)(1/3)}
         {(1/6)(1/6)}
    =
    4.
\end{equation}
Thus, $p$ and $\gamma_p$ have different one-body marginals but belong to
the same relational equivalence class:
\begin{equation}
    [p]=[\gamma_p].
\end{equation}
The table $\gamma_p$ is the canonical representative of this equivalence
class with uniform marginals. For a general bipartite binary table, the uniform-margin representative is
\begin{equation}
    \gamma_p
    =
    \frac{1}{2(1+\sqrt{\omega})}
    \begin{pmatrix}
        \sqrt{\omega} & 1 \\
        1 & \sqrt{\omega}
    \end{pmatrix}.
    \label{eq:2x2-canonical-representative}
\end{equation}
Therefore, knowing $\omega$ uniquely determines the uniform-margin
representative, and conversely.

\section{Multipartite Binary Dependence Nuclei and Complete Coordinates}
\label{sec:complete-coordinates}

\subsection{Review of the log-linear expansion}
\label{subsec:loglinear-review}

With three or more variables, the relational structure no longer contains
only a single degree of freedom. A single odds ratio is no longer sufficient.
The general local reweighting now takes the form
\begin{equation}
    q(x)
    =
    \frac{1}{Z}\,
    p(x)
    \prod_{i=1}^{N} f_i(x_i),
    \qquad
    f_i(x_i)>0.
    \label{eq:general-local-reweighting}
\end{equation}
Products are somewhat inconvenient to handle directly. Taking logarithms
turns them into sums:
\begin{equation}
    \log q(x)-\log p(x)
    =
    -\log Z
    +
    \sum_{i=1}^{N}\log f_i(x_i).
    \label{eq:log-reweighting}
\end{equation}
Since $x_i=\pm1$, any function depending only on $x_i$ can be written as
\begin{equation}
    \log f_i(x_i)
    =
    a_i+b_i x_i.
\end{equation}
This is not an approximation. A function defined on the two points
$\{-1,+1\}$ has only two function values, while $a_i$ and $b_i$ are also
just two numbers:
\begin{equation}
    a_i
    =
    \frac{
        \log f_i(+1)+\log f_i(-1)
    }{2},
\end{equation}
and
\begin{equation}
    b_i
    =
    \frac{
        \log f_i(+1)-\log f_i(-1)
    }{2}.
\end{equation}
Therefore,
\begin{equation}
    \log q(x)-\log p(x)
    =
    c+\sum_i b_i x_i,
    \label{eq:linear-reweighting}
\end{equation}
where all constant terms have been absorbed into $c$.
We thus arrive at an important fact: the only terms that a local positive
reweighting can add to $\log p$ are a constant term and one-body terms.
This suggests a natural idea. If we separate the one-body terms in
$\log p$ from the genuinely multipartite terms, the relational structure
should reveal itself automatically.

Let us first return to the bipartite case and examine the complete expansion
of $\log p$. The four functions $    1$, $ s$, $ t$, $ st$ in \eqref{eq:2x2-reweighting}
take the following values on the four possible outcomes:
\begin{equation}
\begin{array}{c|rrrr}
    (s,t) & 1 & s & t & st \\ \hline
    (+,+) & 1 & 1  & 1  & 1  \\
    (+,-) & 1 & 1  & -1 & -1 \\
    (-,+) & 1 & -1 & 1  & -1 \\
    (-,-) & 1 & -1 & -1 & 1
\end{array}.
\end{equation}
These four sign patterns are linearly independent. Hence any real-valued
function on the four cells can be written uniquely as their linear
combination. In particular, we define
\begin{equation}
    \log p_{st}
    =
    J_{\varnothing}
    +
    J_1 s
    +
    J_2 t
    +
    J_{12}st.
    \label{eq:two-body-loglinear}
\end{equation}
No statistical model has been assumed here. We have merely replaced the
four numbers $    \log p_{++}$, $    \log p_{+-}$, $    \log p_{-+}$, $    \log p_{--}$
by an equivalent set of coordinates.
In particular,
\begin{equation}
    J_{12}
    =
    \frac{1}{4}
    \left(
        \log p_{++}
        +
        \log p_{--}
        -
        \log p_{+-}
        -
        \log p_{-+}
    \right),
\end{equation}
and therefore
\begin{equation}
    J_{12}
    =
    \frac{1}{4}
    \log
    \frac{
        p_{++}p_{--}
    }{
        p_{+-}p_{-+}
    }
    =
    \frac{1}{4}\log\omega.
    \label{eq:J12-odds-ratio}
\end{equation}
Thus, in the bipartite case, $J_{12}$ is simply the logarithmic coordinate
of the odds ratio.

We now introduce the general $N$-partite version of
Eq.~\eqref{eq:two-body-loglinear}. Let
\begin{equation}
    \Omega_N
    =
    \{-1,+1\}^{N},
    \qquad
    p(x)>0
    \quad
    (x\in\Omega_N).
\end{equation}
For every subset $    S\subseteq[N]
    =
    \{1,\ldots,N\}$,
define the product function
\begin{equation}
    \chi_S(x)
    =
    \prod_{i\in S}x_i.
    \label{eq:walsh-character}
\end{equation}
For $S=\varnothing$, we set $    \chi_{\varnothing}(x)=1$.
These functions are also known as Walsh characters. Since $[N]$ has
$2^N$ subsets, there are $2^N$ such functions. They form a complete basis
for real-valued functions on the binary hypercube.\footnote{Indeed, $    2^{-N}
    \sum_{x\in\Omega_N}
    \chi_S(x)\chi_T(x)
    =
    \delta_{ST}$.}
The general form of Eq.~\eqref{eq:two-body-loglinear} is therefore
\begin{equation}
    \log p(x)
    =
    \sum_{S\subseteq[N]}
    J_S(p)\,\chi_S(x).
    \label{eq:walsh-loglinear-expansion}
\end{equation}
More explicitly,
\begin{equation}
    \log p(x)
    =
    J_{\varnothing}
    +
    \sum_i J_i x_i
    +
    \sum_{i<j}J_{ij}x_i x_j
    +
    \sum_{i<j<k}J_{ijk}x_i x_j x_k
    +
    \cdots.
    \label{eq:explicit-loglinear-expansion}
\end{equation}
The coefficients can be extracted directly as
\begin{equation}
    J_S(p)
    =
    2^{-N}
    \sum_{x\in\{\pm1\}^{N}}
    \chi_S(x)\log p(x).
    \label{eq:JS-coefficient}
\end{equation}
The reason Eq.~\eqref{eq:JS-coefficient} isolates a particular $J_S$
is the orthogonality relation\footnote{If $S\neq T$, there is always at least one variable that occurs
in exactly one of the two products. Summing over the two possible values
$+1$ and $-1$ of that variable makes the corresponding contributions
cancel exactly.}
\begin{equation}
    2^{-N}
    \sum_x
    \chi_S(x)\chi_T(x)
    =
    \begin{cases}
        1, & S=T,\\
        0, & S\neq T.
    \end{cases}
    \label{eq:walsh-orthogonality}
\end{equation}

Mathematically, this is the Fourier--Walsh expansion. In the theory of
contingency tables, it is the standard saturated log-linear
expansion~\cite{bishopFienbergHolland1975}.

\subsection{Completeness of the relational data $J_S$}
\label{subsec:completeness-JS}

As already noted, Eq.~\eqref{eq:linear-reweighting} shows that a local
reweighting can change only $    J_{\varnothing}$, $    J_1$, $\ldots$, $J_N$,
while all coefficients carrying at least two indices, $    J_S$, $    |S|\geq2$,
remain invariant. These are precisely the quantities that we expect to
characterize the relational structure.
For example, for three variables,
\begin{align}
    \log p(x_1,x_2,x_3)
    ={}&
    J_{\varnothing}
    +
    J_1x_1
    +
    J_2x_2
    +
    J_3x_3
    \nonumber\\
    &+
    J_{12}x_1x_2
    +
    J_{13}x_1x_3
    +
    J_{23}x_2x_3
    +
    J_{123}x_1x_2x_3.
    \label{eq:three-body-loglinear}
\end{align}
The relational part is therefore expected to be $    J_{12}$, $    J_{13}$, $    J_{23}$, $J_{123}$. In fact, the first three record relations at the pairwise level, while the last
records a genuinely three-body interaction that cannot be absorbed into
pairwise terms.

However, being invariant is not yet the same as being complete. If we had
only proved that the higher-order $J_S$ remain invariant under
reweighting, there could still be further relational information not
captured by them. The key point is that the converse also holds.
Suppose two strictly positive tables $p$ and $q$ satisfy
\begin{equation}
    J_S(p)=J_S(q),
    \qquad
    \forall\,|S|\geq2.
    \label{eq:same-higher-J}
\end{equation}
Then their logarithmic difference contains no multipartite terms:
\begin{equation}
    \log q(x)-\log p(x)
    =
    c+\sum_i b_i x_i.
\end{equation}
Exponentiating,
\begin{equation}
    q(x)
    =
    e^c\,p(x)
    \prod_i e^{b_i x_i}.
\end{equation}
Since $q$ is already normalized, $e^c$ plays precisely the role of
$1/Z$. Hence
\begin{equation}
    q(x)
    =
    \frac{1}{Z}
    p(x)
    \prod_i f_i(x_i),
    \qquad
    f_i(x_i)=e^{b_i x_i}.
\end{equation}
This is exactly a local positive reweighting. We therefore obtain the
following result:\footnote{Historical note. The use of a discrete Fourier expansion of
$\log p$ to define higher-order interactions in multidimensional
contingency tables can be traced at least to
Refs.~\cite{good1963,goodman1964}; in the binary case, this is precisely
a Walsh expansion. In modern language, for example,
Ref.~\cite{buchman2012} explicitly studies binary log-linear models
written as Walsh--Hadamard harmonic expansions. For the purposes of the
present paper, we use this standard structure to organize all Walsh
coefficients with $|S|\geq2$ as complete coordinates of the dependence
nucleus.}

\begin{theorem}[Complete log-linear coordinates of the dependence nucleus]
\label{thm:complete-loglinear-coordinates}
For any strictly positive binary probability tables $p$ and $q$,
\begin{equation}
    p\sim q
    \quad\Longleftrightarrow\quad
    J_S(p)=J_S(q)
    \qquad
    \text{for all }|S|\geq2.
    \label{eq:complete-loglinear-coordinates}
\end{equation}
\end{theorem}
Thus, exactly as in the bipartite case, we may again define an abstract
relational equivalence class $[p]$, while $    \{J_S:\ |S|\geq2\}$
provides a complete set of coordinates for that class.

It is also useful to record the relation to conditional odds ratios.
For simplicity, consider three variables. Fixing the third variable to
$x_3=u$ gives a $2\times2$ slice for $x_1$ and $x_2$. Its conditional
odds ratio is
\begin{equation}
    \omega_{12\mid u}
    =
    \frac{
        p(+,+,u)\,p(-,-,u)
    }{
        p(+,-,u)\,p(-,+,u)
    }.
    \label{eq:conditional-odds-ratio}
\end{equation}
Substituting the log-linear expansion gives
\begin{equation}
    \log\omega_{12\mid u}
    =
    4\left(
        J_{12}+uJ_{123}
    \right).
    \label{eq:conditional-odds-J}
\end{equation}
Conversely,
\begin{equation}
    J_{12}
    =
    \frac{1}{8}
    \left(
        \log\omega_{12\mid+}
        +
        \log\omega_{12\mid-}
    \right),
\end{equation}
and
\begin{equation}
    J_{123}
    =
    \frac{1}{8}
    \left(
        \log\omega_{12\mid+}
        -
        \log\omega_{12\mid-}
    \right).
\end{equation}
Thus, conditional odds ratios are closely tied to the higher-order
$J_S$. 

\section{The Sklar Picture: Can Dependence and Marginals Recover the Whole?}
\label{sec:sklar-reconstruction}

So far, we have defined the relational equivalence class and found a complete
set of coordinates for it. But this still does not reach the full strength of
a Sklar decomposition. A Sklar-type statement requires the following:

\begin{quote}
If an arbitrary relational structure is fixed, and at the same time an
arbitrary collection of one-body marginals is specified, is there exactly one
table in the equivalence class $[p]$ that realizes those marginals?
\end{quote}

\subsection{Proof in the bipartite case}
\label{subsec:sklar-bipartite}

As a warmup, let us first prove the statement in the bipartite case, where
the argument is very simple.

Fix the two marginals
\begin{equation}
    u=\Pr(X=+1),
    \qquad
    v=\Pr(Y=+1).
\end{equation}
The entire table then contains only one remaining free variable, $    z=p_{++}$, 
and can be written as\footnote{Strict positivity requires $    \max\{0,u+v-1\}
    <
    z
    <
    \min\{u,v\}$.}
\begin{equation}
    p(z)
    =
    \begin{pmatrix}
        z & u-z \\
        v-z & 1-u-v+z
    \end{pmatrix}.
    \label{eq:two-body-fixed-marginals}
\end{equation}
The odds ratio is
\begin{equation}
    \omega(z)
    =
    \frac{
        z(1-u-v+z)
    }{
        (u-z)(v-z)
    }.
    \label{eq:omega-z}
\end{equation}
Hence
\begin{equation}
    \frac{d}{dz}\log\omega(z)
    =
    \frac{1}{z}
    +
    \frac{1}{u-z}
    +
    \frac{1}{v-z}
    +
    \frac{1}{1-u-v+z}
    >0.
    \label{eq:omega-monotone}
\end{equation}
Thus, once the marginals are fixed, $\omega(z)$ increases strictly with
$z$. At the two ends of the allowed interval,
\begin{equation}
    \omega(z)\longrightarrow 0
    \qquad\text{and}\qquad
    \omega(z)\longrightarrow\infty,
\end{equation}
respectively. Therefore, every prescribed $    \omega>0$
corresponds to exactly one value of $z$. We have proved
\begin{equation}
    (\text{two marginals},\,\omega)
    \longleftrightarrow
    \text{a unique joint table}.    
    \label{eq:bipartite-sklar}
\end{equation}

To summarize, Sklar's theorem says that if one is given arbitrary marginals
$F_X,F_Y$ and an arbitrary copula $C$ characterizing their dependence,
then one obtains a unique joint distribution $F_{XY}$.
Equation~\eqref{eq:bipartite-sklar} is precisely the concrete realization
of this Sklar decomposition in the strictly positive binary setting:
given arbitrary marginals $    u,v\in(0,1)$,
together with an arbitrary dependence structure $    \omega>0$,
one obtains a unique joint table.

It is useful to comment on the proof strategy, because it will guide the
$N$-partite argument below. Our goal here was: with the marginals $u,v$
fixed, prove that the dependence $\omega$ and the full table $P$ are in
one-to-one correspondence. Directly handling the four entries of $P$ is,
in a sense, ``difficult''---although four entries are of course still
manageable, they are already a miniature version of the $N$-dimensional
problem.
We therefore introduced an internal degree of freedom $z$. Once $u$ and
$v$ are fixed,
\begin{equation}
    P
    \longleftrightarrow
    z
\end{equation}
is manifestly one-to-one. Proving the strict monotonicity of $\omega(z)$
then proves that
\[
    \omega
    \longleftrightarrow
    z
\]
is a bijection. In this way, the logical loop is closed:
\begin{equation}
    \omega
    \longleftrightarrow
    z
    \longleftrightarrow
    P.
\end{equation}
In short, proving a bijection between the dependence $\omega$ and the
internal variable $z$ is equivalent to proving that, once the marginals are
fixed, the dependence and the whole table are in one-to-one correspondence.

\subsection{Proof in the $N$-partite case}
\label{subsec:sklar-Npartite}

\begin{theorem}[Positive binary discrete Sklar decomposition]
\label{thm:positive-binary-sklar}
Fix any strictly positive dependence nucleus $    \mathcal{N}=[p]$.
For every collection of strictly positive one-body marginals, there exists
one and only one joint probability table in $\mathcal{N}$ having those
marginals.
Equivalently,
\begin{equation}
    p
    \longleftrightarrow
    \bigl(
        \text{one-body marginals},
        \mathcal{N}(p)
    \bigr).
    \label{eq:sklar-nucleus-decomposition}
\end{equation}
Using the complete coordinates of Theorem~\ref{thm:complete-loglinear-coordinates},
this can also be written as
\begin{equation}
    \boxed{
    p
    \longleftrightarrow
    \bigl(
        \text{one-body marginals},
        \{J_S:\ |S|\geq2\}
    \bigr).
    }
    \label{eq:sklar-J-decomposition}
\end{equation}
\end{theorem}

We now prove the theorem.

The technical subtlety is that, in the $N$-partite case, if one first fixes
the marginals $    m=(m_1,\ldots,m_N)$,
the remaining relational degrees of freedom are no longer represented by a
single $\omega$, but by $    \{J_S:\ |S|\geq2\}$,
of which there are $    2^N-N-1$. 
We shall therefore adopt a proof strategy complementary to that used in the
bipartite case: we first fix the dependence structure, and then prove that
the marginals and the full joint table are in one-to-one correspondence. First of all, for convenience, we characterize the marginal probabilities of the binary
variables equivalently by their mean values:
\begin{equation}
    m_i
    =
    \mathbb{E}[X_i]
    =
    \Pr(X_i=+1)-\Pr(X_i=-1)
    =
    2r_i-1,
    \label{eq:mi}
\end{equation}
where $    r_i=\Pr(X_i=+1)$.
Clearly, $    r_i\in(0,1)$ and $    m_i\in(-1,1)$ are in
one-to-one correspondence.

Next, define the dependence structure containing all higher-order
coefficients by
\begin{equation}
    R(x)
    =
    \sum_{|S|\geq2}
    J_S\chi_S(x),
    \label{eq:R-def}
\end{equation}
and keep $R(x)$ fixed.
Once this is done, the only nonconstant degrees of freedom left in the Walsh
expansion are the one-body coefficients. Thus every strictly positive
probability table with these same higher-order coefficients has the form
\begin{equation}
    \log p(x)
    =
    c
    +
    R(x)
    +
    \sum_i h_i x_i.
    \label{eq:fixed-R-logp}
\end{equation}
The $h_i$ are one-body natural parameters, but they are not, a priori, the
marginal probabilities themselves.
Writing $    h=(h_1,\ldots,h_N)\in\mathbb{R}^N$,
then, once the dependence structure $R(x)$ is fixed, every strictly
positive table in the same equivalence class $[p]$ can be written uniquely as
\begin{equation}
    p_h(x)
    =
    \frac{
        \exp\!\left[R(x)+h\cdot x\right]
    }{
        Z(h)
    },
    \label{eq:ph}
\end{equation}
where
\begin{equation}
    Z(h)
    =
    \sum_{x\in\Omega_N}
    \exp\!\left[R(x)+h\cdot x\right]
    \label{eq:Zh}
\end{equation}
provides the normalization.
The key point is that, once $R(x)$ is fixed, knowledge of $h$
uniquely determines the whole table $p_h(x)$. In other words,
\begin{equation}
    h
    \longleftrightarrow
    P
\end{equation}
is one-to-one.

We can therefore use $m$ and $h$ as scaffolding for the proof. Our goal is:
with the dependence $R(x)$ fixed, prove that the marginals $m$ and the whole
table $P$ are in one-to-one correspondence. Directly proving ``$    m\longleftrightarrow P
$''
is cumbersome because $P$ contains $2^N$ cells. But
Eq.~\eqref{eq:ph} tells us that ``$    P\longleftrightarrow h$''
is manifestly one-to-one. Hence, if we can prove that ``$    h\longleftrightarrow m$''
is a bijection, the logical loop is closed:
\begin{equation}
    m
    \longleftrightarrow
    h
    \longleftrightarrow
    P.
    \label{eq:logical-loop}
\end{equation}
This proves that, for fixed $R(x)$, the marginals $m$ and the joint table
$P$ are in one-to-one correspondence.

The proof therefore consists of two steps.

\paragraph{Step 1: Injectivity.}

We first prove that the map $    h\longmapsto m(h)$
is injective: two different $h$ can never produce the same $m$.

Define the log-partition function
\begin{equation}
    A(h)
    =
    \log Z(h)
    =
    \log
    \sum_x
    \exp\!\left[
        R(x)+\sum_i h_i x_i
    \right].
    \label{eq:log-partition}
\end{equation}
Differentiating with respect to $h_i$ gives
\begin{equation}
    \frac{\partial A}{\partial h_i}
    =
    \frac{
        \sum_x
        x_i
        \exp\!\left[
            R(x)+\sum_j h_jx_j
        \right]
    }{
        \sum_x
        \exp\!\left[
            R(x)+\sum_j h_jx_j
        \right]
    }.
    \label{eq:A-first-derivative}
\end{equation}
The right-hand side is precisely the expectation value of $X_i$ evaluated
with respect to $p_h$. Therefore the first derivative of $A(h)$ is exactly
$m(h)$:
\begin{equation}
    \nabla A(h)
    =
    m(h).
    \label{eq:mh}
\end{equation}

Differentiating the log-partition function once more gives its Hessian
matrix $    H
    =
    \nabla^2 A(h)$.
Its $(i,j)$ entry is
\begin{equation}
    \frac{\partial^2 A}
         {\partial h_i\partial h_j}
    =
    \operatorname{Cov}_h(X_i,X_j).
    \label{eq:hessian-covariance}
\end{equation}
Here $\operatorname{Cov}$ denotes covariance, while
$\operatorname{Var}$ below denotes variance.\footnote{Explicitly, $    \operatorname{Cov}(X_i,X_j)
    =
    \mathbb{E}
    \left[
        \bigl(X_i-\mathbb{E}[X_i]\bigr)
        \bigl(X_j-\mathbb{E}[X_j]\bigr)
    \right]$,
and $    \operatorname{Var}(X)
    =
    \operatorname{Cov}(X,X)$.}
The key point is that $H$ is positive definite. Indeed, for every nonzero
vector $    v\in\mathbb{R}^N$,
we have
\begin{align}
    v^{T}Hv
    &=
    \sum_i\sum_j
    v_i v_j
    \operatorname{Cov}_h(X_i,X_j)
    \nonumber\\
    &=
    \operatorname{Var}_h
    \left(
        \sum_i v_iX_i
    \right)
    \nonumber\\
    &=
    \operatorname{Var}_h(v\cdot X).
    \label{eq:hessian-positive}
\end{align}
Since $v\neq0$ and $x$ runs over the entire binary hypercube, the linear
combination $v\cdot x$ cannot be constant. A nonconstant random variable
under a strictly positive distribution has strictly positive variance.
Therefore,
\begin{equation}
    \operatorname{Var}_h(v\cdot X)>0.
\end{equation}
Hence the Hessian $    H=\nabla^2A(h) $
is positive definite, and $A(h)$ is strictly convex. An important property
of a differentiable strictly convex function with positive-definite Hessian
is that its gradient is strictly monotone. Using
Eq.~\eqref{eq:mh}, this means that $m(h)$ is strictly monotone; in
particular,
\begin{equation}
    h_1\neq h_2
    \quad\Longrightarrow\quad
    m(h_1)\neq m(h_2).
\end{equation}
Thus the map $h\mapsto m(h)$ is injective.

\paragraph{Step 2: Surjectivity.}

We next prove that $    h\longmapsto m(h)$
is surjective: every admissible $m$ is realized by some $h$.
Given any target vector $    m\in(-1,1)^N$,
we must show that there exists $    h^*\in\mathbb{R}^N$
such that
\begin{equation}
    m(h^*)=m.
    \label{eq:target-m}
\end{equation}
Here $m=(m_1,\ldots,m_N)$ is an $N$-dimensional vector.\footnote{
We shall use the following norm notation:
\begin{itemize}
    \item $\|\cdot\|_2$ is the Euclidean norm, for example $        \|h\|_2
        =
        \sqrt{h_1^2+\cdots+h_N^2}$;

    \item $\|\cdot\|_1$ is the sum of absolute values, for example $        \|u\|_1
        =
        |u_1|+\cdots+|u_N|$;

    \item $\|\cdot\|_\infty$ is the largest absolute value of the
    components of a vector.
\end{itemize}}

By Eq.~\eqref{eq:mh}, solving Eq.~\eqref{eq:target-m} is equivalent to
solving
\begin{equation}
    \nabla A(h)=m.
    \label{eq:gradient-target}
\end{equation}
For this purpose, define the auxiliary function
\begin{equation}
    \Phi_m(h)
    =
    A(h)-m\cdot h.
    \label{eq:Phi-def}
\end{equation}
Its gradient is
\begin{equation}
    \nabla\Phi_m(h)
    =
    \nabla A(h)-m
    =
    m(h)-m.
    \label{eq:Phi-gradient}
\end{equation}
Our task has therefore become the problem of finding the minimum of
$\Phi_m(h)$. The strategy is the following. We first show that
$\Phi_m(h)$ tends to $+\infty$ as $h$ goes to infinity; that is,
$\Phi_m$ is coercive. Its minimum must then lie in a bounded closed region.
A continuous function attains its minimum on a compact set, and because
$\Phi_m$ is strictly convex, this minimum is globally unique. At the
minimum the derivative vanishes, which gives precisely $    \nabla A(h^*)=m$.
This will establish both existence and uniqueness.

Let us now make the argument explicit. We first prove that
$\Phi_m(h)\to+\infty$ as $\|h\|_2\to\infty$.
Write $h$ in terms of its ``length'' and ``direction'':
\begin{equation}
    h=tu,
    \qquad
    t=\|h\|_2,
    \qquad
    \|u\|_2=1.\label{unit}
\end{equation}
Then
\begin{equation}
    \Phi_m(tu)
    =
    A(tu)
    -
    t(u\cdot m).
    \label{eq:Phi-tu}
\end{equation}
By definition,
\begin{equation}
    A(tu)
    =
    \log
    \sum_x
    \exp\!\left[
        R(x)+t\,u\cdot x
    \right].
\end{equation}
Since the binary hypercube is finite, $R(x)$ has a minimum. Denote it by
\begin{equation}
    R_{\min}
    =
    \min_x R(x).
\end{equation}
For $u\cdot x$, because every component of $x$ is either $+1$ or $-1$,
the maximum is obtained by choosing each sign of $x_i$ to agree with the
sign of $u_i$. Therefore,
\begin{equation}
    \max_{x\in\Omega_N}u\cdot x
    =
    \|u\|_1.
\end{equation}
Using these two facts, we immediately obtain
\begin{equation}
    A(tu)
    \geq
    R_{\min}
    +
    t\|u\|_1,
\end{equation}
and hence
\begin{equation}
    \Phi_m(tu)
    \geq
    R_{\min}
    +
    t\bigl(
        \|u\|_1-u\cdot m
    \bigr).
    \label{eq:Phi-lower-bound}
\end{equation}
Now consider the quantity in parentheses. By the triangle inequality,
\begin{equation}
    u\cdot m
    \leq
    |u\cdot m|
    \leq
    \|u\|_1\|m\|_\infty.
\end{equation}
Therefore,
\begin{equation}
    \|u\|_1-u\cdot m
    \geq
    \|u\|_1
    \bigl(
        1-\|m\|_\infty
    \bigr).
    \label{eq:positive-bracket}
\end{equation}
Since $    m\in(-1,1)^N$,
we have $    \|m\|_\infty<1$.
And because of \eqref{unit}, $    \|u\|_1\geq1$.
Thus
\begin{equation}
    \|u\|_1-u\cdot m
    \geq
    1-\|m\|_\infty
    >0.
\end{equation}
Consequently,
\begin{equation}
    \Phi_m(tu)
    \geq
    R_{\min}
    +
    t\bigl(1-\|m\|_\infty\bigr)
    \longrightarrow
    +\infty
    \qquad
    (t\to\infty).
    \label{eq:Phi-coercive}
\end{equation}
Hence $\Phi_m$ is coercive.
It follows that its minimum must lie in a finite bounded region. To make
this explicit, choose an arbitrary reference point $h_0$ and define
\begin{equation}
    K
    =
    \left\{
        h\in\mathbb{R}^N:
        \Phi_m(h)\leq\Phi_m(h_0)
    \right\}.
\end{equation}
Because $\Phi_m(h)\to+\infty$ at infinity, $K$ is bounded. Since
$\Phi_m$ is continuous, $K$ is also closed. Hence $K$ is compact.
A continuous function attains its minimum on a compact set, so
$\Phi_m$ has a minimum in $K$. Outside $K$, every value of $\Phi_m$ is
larger than $\Phi_m(h_0)$, so this minimum is also a global minimum.

Furthermore, since $A(h)$ is strictly convex and $m\cdot h$ is linear, $    \Phi_m(h)=A(h)-m\cdot h$ is also strictly convex. Its global minimum is therefore unique. Denote the
minimizer by $h^*$.
Because $h^*$ lies in the unconstrained space $\mathbb{R}^N$ and
$\Phi_m$ is differentiable, Fermat's theorem gives
\begin{equation}
    \nabla\Phi_m(h^*)=0.
\end{equation}
Using Eq.~\eqref{eq:Phi-gradient}, we immediately obtain
\begin{equation}
    \nabla A(h^*)=m.
\end{equation}
Finally, by Eq.~\eqref{eq:mh},
\begin{equation}
    m(h^*)=m.
\end{equation}
Thus, for every admissible target $m$, the unique minimizer of
$\Phi_m(h)$ gives the corresponding $h^*$. Surjectivity is proved.

Combining the two steps, injectivity guarantees ``at most one table'':
two different $h$ cannot produce the same $m$. Surjectivity guarantees
``at least one table'': every admissible $m$ is produced by some $h$.
Together they establish a bijection $    h
    \longleftrightarrow
    m$. Since, for fixed $R(x)$, $    h
    \longleftrightarrow
    P$
is already one-to-one, we conclude that $    m
    \longleftrightarrow
    P$
is also one-to-one.
This completes the proof: once the higher-order dependence structure is
fixed, every admissible collection of one-body marginals corresponds to
one and only one joint probability table.

\section{Discussion and Conclusion}
\label{sec:discussion}

\subsection{A statistical-mechanical interpretation}
\label{subsec:gibbs-interpretation}

We now show that the framework developed here for positive binary
probability tables admits a rather direct interpretation in statistical
mechanics: the higher-order coefficients $J_S$ correspond, in finite-spin
Gibbs systems, precisely to two-body and higher-order interaction couplings.

To see this, let us begin with the most familiar Ising model. Consider two
classical spins $    s_1,s_2\in\{-1,+1\}$.
The simplest Ising Hamiltonian is
\begin{equation}
    H(s_1,s_2)
    =
    -K s_1s_2
    -
    B_1s_1
    -
    B_2s_2.
    \label{eq:two-spin-Hamiltonian}
\end{equation}
Here $K$ is the coupling strength between the two spins, while $B_1$ and
$B_2$ are local external-field parameters at the two sites, with factors
such as the magnetic moment absorbed into their definitions.
At temperature $T$, the corresponding Gibbs distribution is
\begin{equation}
    p(s_1,s_2)
    =
    \frac{1}{Z}
    \exp\!\left[
        \beta K s_1s_2
        +
        \beta B_1s_1
        +
        \beta B_2s_2
    \right],
    \label{eq:two-spin-Gibbs}
\end{equation}
where $    \beta
    =
    \frac{1}{k_{\mathrm B}T}$.
Now compare this with our log-linear expansion,
\begin{equation}
    p(s_1,s_2)
    =
    \frac{1}{Z}
    \exp\!\left[
        J_{12}s_1s_2
        +
        h_1s_1
        +
        h_2s_2
    \right].
\end{equation}
Comparing the two expressions term by term immediately gives
\begin{equation}
    J_{12}=\beta K,
    \qquad
    h_i=\beta B_i.
    \label{eq:ising-dictionary}
\end{equation}
Thus, in this concrete physical model, $J_{12}$ corresponds directly to
the spin--spin coupling strength.

More generally, for an $N$-spin system, consider a Hamiltonian of the form
\begin{equation}
    H(x)
    =
    -
    \sum_{|S|\geq2}
    K_S
    \prod_{i\in S}x_i
    -
    \sum_i B_i x_i.
    \label{eq:general-spin-Hamiltonian}
\end{equation}
At fixed temperature $T$, its Gibbs equilibrium distribution is
\begin{equation}
    p(x)
    =
    \frac{1}{Z}
    \exp\!\left[
        \beta
        \sum_{|S|\geq2}
        K_S
        \prod_{i\in S}x_i
        +
        \beta
        \sum_i B_i x_i
    \right].
    \label{eq:general-spin-Gibbs}
\end{equation}
This is exactly of the form of Eq.~\eqref{eq:ph}, under the identification
\begin{equation}
    J_S=\beta K_S,
    \qquad
    h_i=\beta B_i.
    \label{eq:gibbs-dictionary}
\end{equation}
Hence the coefficients $J_S$, with $|S|\geq2$, correspond to many-body
interaction couplings, while the $h_i$ also acquire a direct physical
meaning: they are the local external-field parameters.
Let us describe this correspondence a little more carefully. Consider the
$i$th spin, $    X_i\in\{-1,+1\}$.
Its marginal distribution consists of the two probabilities $    p_i(+1)$, $    p_i(-1)$,
whose sum is one. Thus this marginal distribution contains only one
independent degree of freedom. Now, the local magnetization is precisely the expectation value of $X_i$:
\begin{equation}
    m_i
    =
    \langle X_i\rangle
    =
    (+1)p_i(+1)+(-1)p_i(-1).
    \label{eq:local-magnetization}
\end{equation}
This is exactly Eq.~\eqref{eq:mi}. Therefore, under the present physical
interpretation, the $m_i$ introduced in the previous section are precisely
the local magnetizations. Specifying $m_i$ is equivalent to specifying the
complete marginal distribution of the $i$th spin, and conversely.
Theorem~\ref{thm:positive-binary-sklar}, which characterizes the discrete
Sklar decomposition in probabilistic language, therefore acquires the
following interpretation in statistical mechanics:

\begin{quote}
Given any finite collection of many-body coupling parameters $J$ and any
local magnetization configuration $m$, there exists a unique Gibbs
equilibrium distribution $p$ realizing them.
\end{quote}

The proof of Theorem~\ref{thm:positive-binary-sklar} in
Sec.~\ref{subsec:sklar-Npartite} itself also admits a statistical-mechanical
interpretation. Recall that, after fixing all many-body couplings $J_S$, we
proved that $    h
    \longleftrightarrow
    m$
is a bijection. Since $h$ corresponds to the local external fields, this
bijection can be read as follows:

\begin{quote}
For any fixed finite many-body interaction structure, and for any target
local magnetization $    m\in(-1,1)^N$,
there exists one and only one collection of local external fields $h$ for
which the Gibbs distribution realizes precisely those magnetizations.
\end{quote}

In other words, the interactions determine which external fields are
required to realize a given set of marginals, but they do not restrict
which strictly positive marginals can be realized.

This statistical-mechanical interpretation also gives the canonical
uniform-margin representative $\gamma_p$ a simple physical meaning.
For a binary variable, a uniform marginal is equivalent to $    m_i=0$.
We therefore obtain, almost for free, a particular consequence: for fixed
many-body couplings, there exists a unique collection of external fields
$h^\star$ for which all local magnetizations vanish simultaneously.
Of course,
\begin{equation}
    m_i=0
    \qquad\not\Rightarrow\qquad
    h_i=0
\end{equation}
in general, and the two notions should not be confused.

Finally, it is important to note that fixing the $J_S$ does not mean that
correlation functions in the usual sense remain unchanged. As the local
external fields are varied, $    \langle X_iX_j\rangle$
and the corresponding covariances will in general change. In this sense,
the dependence nucleus is closer to the \emph{interaction structure} of a
Gibbs model than to a correlation function observed in one particular
state.

\subsection{Conclusion}
\label{subsec:conclusion}

The ambition of this paper has not been to develop a general theory of
discrete Sklar structures for finite multivariate statistics. Rather, by
restricting ourselves to positive binary probability distributions, we have
tried to display, transparently and explicitly, a discrete Sklar structure
that appears in a more abstract form in the general theory.
Starting from a probability table $    p(x_1,\ldots,x_N)$,
we sought to separate its information into two parts: 
    \text{statistics of the individual parties}
and
    \text{relations among the parties}.
This separation was not simply declared on intuitive grounds. We needed a
definite mathematical operation under which what we call the ``relation''
could still be identified while the statistical laws of the individual
parties were changed. This operation is local reweighting. It allows us to
say that two probability tables, even with different marginals, may belong
to the same dependence nucleus.
We then showed transparently
 that these two pieces of information can be specified
independently: for any dependence nucleus and any collection of strictly
positive one-body marginals, there exists a unique strictly positive joint
distribution realizing both. Thus, the distinction is not merely conceptual;
the two components can be varied independently and recombined uniquely.
This is the real content of the Sklar idea in the present concrete model:
it gives a particularly explicit way to see how ``the statistical behaviors of
the individual parties'' and ``the relations among the parties'' can be
described separately, and how they can then be uniquely recombined into a
whole.
One point, however, should be emphasized. Our definition of the nucleus is
an equivalence class under a particular transformation group. It captures a
well-defined notion of relational invariance, but we do not claim, without
further argument, that it is the unique mathematical definition of every
possible notion of relation.

We have also shown that fixed many-body couplings in statistical mechanics
provide a concrete realization of this description.  Conceptually, this statistical-mechanical interpretation does not go beyond
our original Sklar idea. Rather, the Sklar idea allows us to understand anew,
from a different angle, the familiar roles of external fields and interactions
in statistical mechanics.
Nevertheless, this physical interpretation does provide a useful piece of
support for the picture developed here. The dependence nucleus, which was
defined abstractly, corresponds in a concrete and mature physical theory to
an interaction structure that can be identified explicitly.
It also reminds us of a simple point: the relational structure of a whole
need not be identified with the statistical correlations displayed by that
whole under one particular set of conditions. In a Gibbs system with fixed
couplings and varying external fields, this distinction is completely
explicit.

Finally, the organization developed in this paper originally arose from a
problem in quantum-state reconstruction~\cite{Lin:2026meu}. There one needs
to separate one-body statistics from multipartite dependence in each
multipartite Born probability table. Here we have extracted the purely
classical part of that problem and studied it as an independent problem in
discrete probability. Nothing in the present analysis depends on quantum
mechanics.

\section*{Acknowledgments}
 This work is supported by the NSFC Grant No.1250050230.


\bibliographystyle{unsrt}
\bibliography{references}

@article{sklar1959,
  author  = {Sklar, Abe},
  title   = {Fonctions de r{\'e}partition {\`a} n dimensions et leurs marges},
  journal = {Publications de l'Institut de Statistique de l'Universit{\'e} de Paris},
  volume  = {8},
  pages   = {229--231},
  year    = {1959}
}

@article{geenens2020,
  author  = {Geenens, Gery},
  title   = {Copula modeling for discrete random vectors},
  journal = {Dependence Modeling},
  volume  = {8},
  number  = {1},
  pages   = {417--440},
  year    = {2020},
  doi     = {10.1515/demo-2020-0022}
}

@misc{geenens2025,
  author = {Geenens, Gery and Kojadinovic, Ivan and Martini, Tommaso},
  title  = {The empirical discrete copula process},
  year   = {2025},
  note   = {\href{https://arxiv.org/abs/2506.12316}{arXiv:2506.12316 [math.ST]}}
}

@article{yule1900,
  author  = {Yule, G. Udny},
  title   = {On the Association of Attributes in Statistics:
             With Illustrations from the Material of the Childhood Society, \&c.},
  journal = {Philosophical Transactions of the Royal Society of London. Series A},
  volume  = {194},
  pages   = {257--319},
  year    = {1900},
  doi     = {10.1098/rsta.1900.0019}
}

@article{yule1912,
  author  = {Yule, G. Udny},
  title   = {On the Methods of Measuring Association between Two Attributes},
  journal = {Journal of the Royal Statistical Society},
  volume  = {75},
  number  = {6},
  pages   = {579--642},
  year    = {1912},
  doi     = {10.1111/j.2397-2335.1912.tb00463.x}
}

@article{goodmanKruskal1954,
  author  = {Goodman, Leo A. and Kruskal, William H.},
  title   = {Measures of Association for Cross Classifications},
  journal = {Journal of the American Statistical Association},
  volume  = {49},
  number  = {268},
  pages   = {732--764},
  year    = {1954},
  doi     = {10.1080/01621459.1954.10501231}
}

@article{mosteller1968,
  author  = {Mosteller, Frederick},
  title   = {Association and Estimation in Contingency Tables},
  journal = {Journal of the American Statistical Association},
  volume  = {63},
  number  = {321},
  pages   = {1--28},
  year    = {1968},
  doi     = {10.1080/01621459.1968.11009219}
}

@incollection{bergsmaRudas2002Variation,
  author    = {Bergsma, Wicher P. and Rudas, Tam{\'a}s},
  title     = {Variation Independent Parameterizations of Multivariate Categorical Distributions},
  booktitle = {Distributions with Given Marginals and Statistical Modelling},
  editor    = {Cuadras, C. M. and Fortiana, J. and Rodr{\'i}guez-Lallena, J. A.},
  publisher = {Kluwer Academic Publishers},
  address   = {Dordrecht},
  pages     = {21--27},
  year      = {2002},
  doi       = {10.1007/978-94-017-0061-0_3}
}

@book{rudas2018,
  author    = {Rudas, Tam{\'a}s},
  title     = {Lectures on Categorical Data Analysis},
  series    = {Springer Texts in Statistics},
  publisher = {Springer},
  address   = {New York},
  year      = {2018},
  doi       = {10.1007/978-1-4939-7693-5}
}

@inproceedings{liu2025DiscreteCopulaDiffusion,
  author    = {Liu, Anji and Broadrick, Oliver and Niepert, Mathias
               and Van den Broeck, Guy},
  title     = {Discrete Copula Diffusion},
  booktitle = {International Conference on Learning Representations},
  year      = {2025},
  eprint    = {2410.01949},
  archivePrefix = {arXiv},
  primaryClass  = {cs.LG}
}

@article{good1963,
  author  = {Good, I. J.},
  title   = {Maximum Entropy for Hypothesis Formulation, Especially for Multidimensional Contingency Tables},
  journal = {The Annals of Mathematical Statistics},
  volume  = {34},
  number  = {3},
  pages   = {911--934},
  year    = {1963},
  doi     = {10.1214/aoms/1177704014}
}

@article{goodman1964,
  author  = {Goodman, Leo A.},
  title   = {Interactions in Multidimensional Contingency Tables},
  journal = {The Annals of Mathematical Statistics},
  volume  = {35},
  number  = {2},
  pages   = {632--646},
  year    = {1964},
  doi     = {10.1214/aoms/1177703561}
}

@inproceedings{buchman2012,
  author    = {Buchman, David and Schmidt, Mark and Mohamed, Shakir
               and Poole, David and de Freitas, Nando},
  title     = {On Sparse, Spectral and Other Parameterizations of Binary Probabilistic Models},
  booktitle = {Proceedings of the Fifteenth International Conference on Artificial Intelligence and Statistics},
  series    = {Proceedings of Machine Learning Research},
  volume    = {22},
  pages     = {173--181},
  year      = {2012},
  publisher = {PMLR}
}

@book{bishopFienbergHolland1975,
  author    = {Bishop, Yvonne M. M. and Fienberg, Stephen E. and Holland, Paul W.},
  title     = {Discrete Multivariate Analysis: Theory and Practice},
  publisher = {MIT Press},
  address   = {Cambridge, MA},
  year      = {1975}
}

@misc{Lin:2026meu,
  author = {Lin, Yi-Yu},
  title  = {Sklar's Theorem and Quantum State Reconstruction from All-Context Dependence},
  year   = {2026},
  note   = {\href{https://arxiv.org/abs/2609.08035}{arXiv:2609.08035 [quant-ph]}}
}

\end{document}